\documentclass[sigconf]{acmart}

\newcommand \ignore[1]{}
\usepackage{graphics}
\usepackage{caption}
\usepackage{subcaption}
\usepackage{color, colortbl}
\definecolor{LightCyan}{rgb}{0.88,1,1}
\definecolor{Gray}{gray}{0.9}
\definecolor{mycolor}{rgb}{1,0,1} %{.5,0,0} %
\usepackage{nameref}
\usepackage{cases}

\AtBeginDocument{%
  \providecommand\BibTeX{{%
    \normalfont B\kern-0.5em{\scshape i\kern-0.25em b}\kern-0.8em\TeX}}}
\setcopyright{acmcopyright}
\copyrightyear{2020}
\acmYear{2020}
\acmConference[PervasiveHealth]{Pervasive Health}{2020}{Atlanta, GA}
\acmBooktitle{PervasiveHealth, 2020, Atlanta GA}
\acmPrice{15.00}
\acmISBN{978-1-4503-XXXX-X/20/09}

\begin{document}

\title{Adherence to Personal Health Devices: A Case Study in Diabetes Management}

%% Author List

\author{Sudip Vhaduri}
\affiliation{
  \institution{Fordham University}
  \streetaddress{Address}
  \city{Bronx}
  \state{NY}
  \postcode{10458}
}
\authornote{Both authors contributed equally to this research.}

\author{Temiloluwa Prioleau}
\affiliation{
  \institution{Dartmouth College}
  \streetaddress{Address}
  \city{Hanover}
  \state{NH}
  \postcode{Zipcode}
}
\authornotemark[1]

\renewcommand{\shortauthors}{Vhaduri and Prioleau}

\begin{abstract}
Personal health devices can enable continuous monitoring of health parameters. However, the benefit of these devices is often directly related to the frequency of use. Therefore, adherence to personal health devices is critical. This paper takes a data mining approach to study continuous glucose monitor use in diabetes management. We evaluate two independent datasets from a total of 44 subjects for 60 - 270 days. Our results show that: 1) missed target goals (i.e. suboptimal outcomes) is a factor that is associated with wearing behavior of personal health devices, and 2) longer duration of non-adherence, identified through missing data or data gaps, is significantly associated with poorer outcomes. More specifically, we found that up to 33\% of data gaps occurred when users were in abnormal blood glucose categories. The longest data gaps occurred in the most severe (i.e. \textit{very low / very high}) glucose categories. Additionally, subjects with poorly-controlled diabetes had longer average data gap duration than subjects with well-controlled diabetes. This work contributes to the literature on the design of context-aware systems that can leverage data-driven approaches to understand factors that influence non-wearing behavior. The results can also support targeted interventions to improve health outcomes.

%Personal health devices can enable continuous monitoring of health parameters. However, benefit is often directly related to frequency of use. Therefore, adherence is critical. This paper presents a  case study of continuous glucose monitor use in diabetes management. We evaluate two independent datasets from a total of 44 subjects for 60 - 270 days. Our results show that: 1) suboptimal outcomes is a factor that affects wearing behavior of personal health devices, and 2) longer durations of non-wearing (i.e. data gaps) is significantly associated with poorer performance. More specifically, we found that up to 33\% of data gaps occurred when users were in abnormal blood glucose categories. Additionally, the longest data gaps occurred in the most severe (i.e. \textit{very low / very high}) glucose categories. In response, we recommend adherence analysis to evaluate data-driven factors that influence non-wearing behavior. This can support targeted interventions to improve health outcomes. 

\end{abstract}

\begin{comment}
\category{}{Human-centered computing}{Ubiquitous and mobile devices} \category{}{Applied Computing}{Health informatics}

    \category{H.5.m.}{Information Interfaces and Presentation
  (e.g. HCI)}{Miscellaneous} \category{See
  \url{http://acm.org/about/class/1998/} for the full list of ACM
  classifiers. This section is required.}{}{}
\end{comment}

\begin{CCSXML}
<ccs2012>
<concept>
<concept_id>10003120.10003121.10003122.10003334</concept_id>
<concept_desc>Human-centered computing~User studies</concept_desc>
<concept_significance>500</concept_significance>
</concept>
<concept>
<concept_id>10003120.10003121.10011748</concept_id>
<concept_desc>Human-centered computing~Empirical studies in HCI</concept_desc>
<concept_significance>300</concept_significance>
</concept>
</ccs2012>
\end{CCSXML}

\ccsdesc[500]{Human-centered computing~User studies}
\ccsdesc[300]{Human-centered computing~Empirical studies in HCI}

% Author Keywords
\keywords{Continuous glucose monitors, mobile health, personal informatics, wearable systems}

%%
%% This command processes the author and affiliation and title information and builds the first part of the formatted document.
\maketitle
%%%%%%%%%%%%%%%%%%%%%%%%%%%%%%%%%%%%%%%%%%%%%%%%%%%
%% Sections
%%%%%%%%%%%%%%%%%%%%%%%%%%%%%%%%%%%%%%%%%%%%%%%%%%%
% Add reference: Diabetes device use in adults with type 1 diabetes: Barriers to uptake and potential intervention targets, also look at https://hcp.eversensediabetes.com/why-eversense-cgm for more on weartime

\section{Introduction}\label{introduction}
Personal health devices (PHD), often in the form of mobile and wearable systems, are particularly useful for pervasive monitoring of health status and vital signs \cite{appelboom2014smart,chan2012smart}. These technologies provide unique opportunities for early diagnosis of diseases, management of chronic conditions, and prompt-response to emergency situations \cite{bonato2010advances}. PHDs have been employed for monitoring of many conditions such as heart disease \cite{oresko2010wearable}, Parkinson's disease \cite{rigas2012assessment}, and diabetes \cite{desalvo2013continuous}. Despite, the potential advantages of these technologies, the benefit is often proportional to the frequency of use \cite{rodbard2016continuous, makela2013adherence}. For example, the American Diabetes Association (ADA) states that frequency of PHD use, specifically continuous glucose monitors, is the "greatest predictor" for lowering hemoglobin A1C - a primary clinical outcome for diabetes management \cite{ADA2017}. Therefore, the notion of adherence to PHDs is critical. A person who uses these devices as intended often achieves better outcomes. Conversely, a person who does not use these devices as intended often achieves suboptimal outcomes. However, it is important to note that wearable PHDs are facilitators, and not drivers, of health behavior change \cite{patel2015wearable}.

There are several definitions of adherence \cite{tang2018adherence}. However, in this paper, we adopt the definition that adherence means complying to a recommended regimen to achieve the best outcome. A recommended regimen can be in the form of guidelines such as taking 10,000 steps per day or prescriptive such as monitoring blood glucose before and after meals. Given the rise of commercial wearable devices in today's society, recent work has focused on adherence to non-prescription PHDs such as physical activity trackers \cite{jeong2017smartwatch,tang2018adherence,doherty2017large}. However, adherence to prescription PHDs such as inhalers for asthma control \cite{makela2013adherence} or continuous glucose monitors used in diabetes management \cite{giani2017continuous} is arguably more important. In the case of asthma or diabetes, there can be an immediate risk or an undesired health event associated with non-adherence. 

This paper focuses on a case study of adherence to PHDs in diabetes care for two key reasons. Firstly, diabetes is the 7$^{th}$ leading cause of death and it affects up to 9.4\% of people in the U.S. \cite{CDC_Diabetes2017}.  This is a significant fraction of the population. Secondly, and equally as important, PHDs in diabetes management are relatively advanced as there exists wearable devices for continuous monitoring of the most relevant biomarker (i.e. blood glucose) \cite{desalvo2013continuous,rodbard2016continuous}. Similar devices for management of other chronic conditions (e.g. heart disease, mental illness, and obesity) are lagging. However, extensive effort is being committed to develop wearable alternatives for continuous 24-hour monitoring of relevant biological and behavioral markers \cite{chan2012smart,pantelopoulos2010survey,prioleau2017unobtrusive}. We envision that findings from this study on diabetes can inform PHD data analysis in other domains. 

\begin{figure}
    \centering
    \includegraphics[width=0.5\columnwidth]{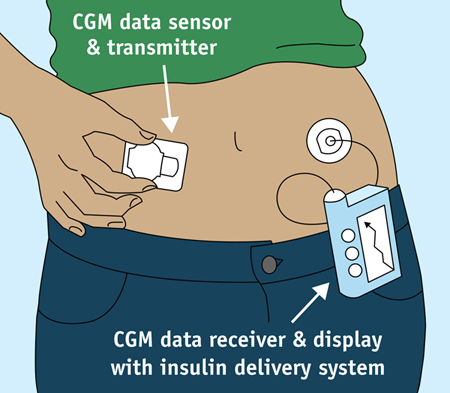}
    \caption{Personal health devices for diabetes care: A continuous glucose monitor (left), blood glucose display and insulin pump (right)~\cite{NIDDK_CGM}.}
    \label{fig:CGM}
\end{figure}

A revolutionary innovation in diabetes care was the development of a continuous glucose monitor (CGM). As shown in Figure \ref{fig:CGM}, it is a minimally-invasive wearable device that enables real-time monitoring of blood glucose (BG) levels from sampling concentrations in the interstitial fluid \cite{desalvo2013continuous}. In comparison to intermittent self-monitoring using glucose meters, CGMs enable the ability to dynamically adapt management strategies such as food intake, exercise, and medication-use to real-time glucose trends. Proper use of CGMs has been shown to reduce risk factors of diabetes such as severe low blood glucose and micro-/macro-vascular complications \cite{danne2017international, desalvo2013continuous,ritholz2010psychosocial,rodbard2016continuous,schmidt2012psychosocial}. However, as is the case with any wearable PHDs, people do not always use them as recommended \cite{giani2017continuous,ritholz2010psychosocial, schmidt2012psychosocial}. 

The objective of this paper is to assess factors that affect adherence (i.e. wearing or use behavior) of personal health devices. More specifically, we seek to investigate: "whether and to what extent achieving target glycemic goals affect wearing behavior of continuous glucose monitors used in diabetes management." Based on data from larger project, %called \textit{Digital SMD - Developing Digital Tools for Improved Self-Management of Diabetes}
we evaluated 60 - 270 days of CGM data from 44 subjects with diabetes and found that:
\begin{enumerate}
    \item Performance toward target goal and age are two factors that influence adherence to PHD.
    \item Longer duration of data gaps occurred in suboptimal BG categories and the longest data gaps occurred in the most severe (i.e. \textit{very low/very high}) BG categories.
    \item Subjects with poorly-controlled diabetes had on average longer data gap durations, indicative of worse adherence, than subjects with well-controlled diabetes.
    \item Older subjects (age: > 40-yrs) had significantly worse adherence to PHDs, evident by longer data gap durations, compared to younger subjects (age: 24 - 40-yrs).
    \item PHD adherence varied across individuals and showed to be subject-dependent.
    
\end{enumerate}

A key recommendation from this work is for development of context-aware PHDs that implement data-driven adherence analysis in embedded algorithms to improve wearing behavior, guide interventions, and positively affect health outcomes. In the case of CGM use in diabetes management, non-wearing behavior influenced by suboptimal BG can be identified based on the BG category users were in prior to the start of data gaps (or missing data events). Adherence analysis to PHDs is important in many health applications \cite{johnson2016methods}. Therefore, we expect results from this work to inform research in other domains.  However, a potential limitation of this work is the assumption that data gaps or missing data is directly indicative of non-adherence to PHDs, specifically CGMs in this study. 

\begin{comment}
Our results show that performance toward the target goal is a factor that influences non-wearing behavior. More specifically, we found that up to 33\% of data gaps occurred when users were in abnormal BG categories. The longest data gaps occurred in the most severe (i.e. \textit{very low/very high}) BG categories. Additionally, subjects with poorly-controlled diabetes had longer average durations of missing data compared to subjects with well-controlled diabetes. 
\end{comment}

% Reference: Mobile health devices: will patients actually use them?

\section{Related Work}\label{relatedwork}
This section reviews relevant literature on personal health data and interpretation with a focus on non-prescription and prescription PHDs.

\subsection{Adherence to Non-prescription PHDs}
Adherence to PHDs has more commonly been studied for consumer wearable systems such as physical activity trackers and smartwatches \cite{epstein2016abandonment,jeong2017smartwatch,tang2018adherence,vhaduri2019multi, meyer2017identification,liu2017characterizing,doherty2017large}. Jeong et al. evaluated smartwatch use amongst 50 college students to understand factors that affect wearing behavior \cite{jeong2017smartwatch}. They found that participants wore their smartwatches for an average of 10.9 and 8.4 hours/day on weekdays and weekends, respectively. Users of such wearable devices were classified into three categories, namely, work-hour wearers, day-time wearers, and all-day wearers. However, only a small percentage of users (about 10\%) are all-day wearers, most users tend to take off their device before bed-time \cite{jeong2017smartwatch,lyons2015dumbwatch}. Tang and Kay \cite{tang2017harnessing} studied adherence to long-term FitBit users. They showed that users benefited from a calendar-view display of daily and hourly adherence in association with the adherence goal. As expected, users cannot achieve the optimal benefit from PHDs without wearing the device. A large scale population study by Doherty et al. \cite{doherty2017large} found that age and time of day are key variables associated with compliance to physical activity trackers. Additionally, several studies have shown that there is high abandonment of consumer wearable devices after about 2-months \cite{clawson2015no,lazar2015abandon,shih2015adoption}. Some reasons for abandonment include devices not fitting with user's conceptions of themselves, discomfort with information revealed, and the collected data not being perceived as helpful for continued use \cite{epstein2016abandonment,lazar2015abandon}. These findings are applicable for leisurely-used, non-prescription PHDs, however, they do not exactly translate to prescription PHDs needed for management of a health condition. 

%Vhaduri and Poellabauer \cite{vhaduri2019multi} studied 400 Fitbit users based for 17-months and reported an average wear time of 20 hours/day ($\pm$ 1.5 hours/day). 

\subsection{Adherence to Prescription PHDs}
In a review on adherence to inhaler devices, non-adherence was found to be influenced by patient knowledge/education, convenience of the device, age, adverse effects, and associated costs \cite{makela2013adherence}. Likewise some factors that have been identified which limit adherence to CGM devices used in diabetes management include cost, sensor discomfort, device inaccuracy, and general usability issues \cite{desalvo2013continuous, rodbard2016continuous,schmidt2012psychosocial,petrie2017improving}. A 6-month clinical trial found that the mean CGM adherence in patients with type 1 diabetes differed across age groups with the highest adherence found in adults (ages: > 18 years) and lowest adherence found in adolescents (ages: 12 - 18 years). \cite{giani2017continuous}. Other studies have found that psychosocial factors such as coping skills, body image, and support from loved ones are associated with the use of CGMs \cite{ritholz2010psychosocial,schmidt2012psychosocial}. The aforementioned studies highlight demographic, usability, cost, and psychosocial factors that influence accumulative adherence, however, little effort has focused on understanding contextual factors that affect day-to-day adherence. The recent papers by Raj et al. \cite{raj2019clinical,raj2019BGHigher} highlight the importance of evaluating clinical data from PHDs in context. More specifically, they show that management of chronic conditions such as diabetes varies in different contextual settings influenced by time, location, people, and emotional state. Unlike the aforementioned work, this paper takes a quantitative, data-driven approach to investigate whether and to what extent management outcomes influence adherence to PHDs. This insight can inform the design of context-aware algorithms that include adherence analysis to identify subject-specific factors associated with non-wearing behavior, provide target interventions, and improve outcomes.

\section{Background} \label{background}
Diabetes is characterized by impaired glucose metabolism. Therefore, a person with diabetes should be constantly aware of the many factors that can affect their body's glucose levels including food, activity, medication, environment, and behaviors in daily living \cite{Brown_42FactorsBG}. The primary management goal is to minimize the occurrence of hypoglycemic (i.e. \textit{low} BG) and hyperglycemic (i.e. \textit{high} BG) events \cite{desalvo2013continuous,rodbard2016continuous,petrie2017improving}. Based on clinical research \cite{danne2017international}, there are five BG categories that are important, namely:
\begin{enumerate}
    \item \textit{\textbf{Very Low}}: Periods of BG readings < 54 mg/dL. This is considered a clinically-significant hypoglycemic event that may require immediate action.
    \item \textit{\textbf{Low}}: Periods of BG readings between 54 - 70 mg/dL. It is recommended to set a CGM hypoglycemia alert for this category to reduce the risk of a more severe event.
    \item \textit{\textbf{Normal}}: Periods of BG readings between 70 - 180 mg/dL. This is considered the target range and the goal is to maximize time spent in this range. 
    \item \textit{\textbf{High}}: Periods of BG readings between 180 - 250 mg/dL. It is recommended to set a CGM hyperglycemia alert for this category to reduce the risk of a more severe event.
    \item \textit{\textbf{Very High}}: Periods of BG readings > 250 mg/dL. This is considered a clinically-significant hyperglycemic event that may require immediate action.
\end{enumerate}

In this work, we use the above categorization of BG readings to evaluate adherence and wearing behavior of CGMs amongst persons with diabetes. It is important to note that CGMs are not perfect and can have inaccuracies in the range of +/- 10\% \cite{rodbard2016continuous,desalvo2013continuous}. Additionally, majority of these devices need to be calibrated using conventional finger-prick method and a blood glucose meter \cite{danne2017international}. Nonetheless, CGMs are the gold standard PHD for real-time monitoring of BG in diabetes \cite{vashist2013continuous}, therefore, they were used in this study.
\section{Data Description and Pre-processing} \label{dataset}
\begin{figure*}
\centering
\begin{subfigure}{0.5\textwidth}
  \centering
  \includegraphics[width=1\linewidth]{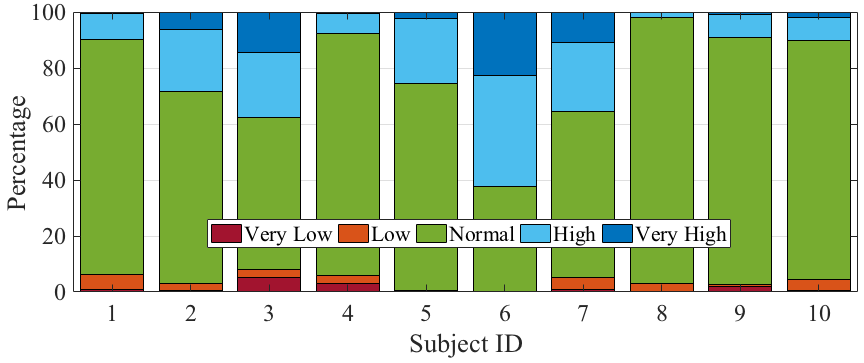}
  \caption{Dataset-1}
  \label{fig:subjBGCounts_5Cat_ds1}
\end{subfigure}%
%\newline
%\newline
%\newline
%\newline
\begin{subfigure}{0.5\textwidth}
  \centering
  \includegraphics[width=1\linewidth]{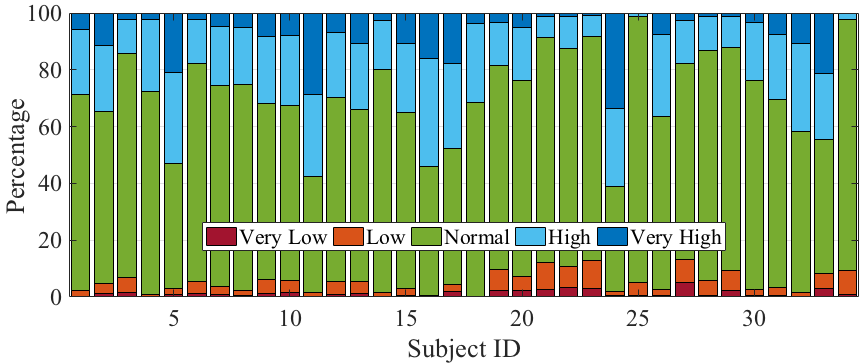}
  \caption{Dataset-2}
  \label{fig:subjBGCounts_5Cat_ds2}
\end{subfigure}%
\caption{Stacked bar plots showing subject-level sample distribution across the 5 key BG categories.}
\label{fig:subjBGCounts_5Cat}
\end{figure*}

All the data used in this study was contributed to the %\textit{Digital SMD} 
research project by members of online diabetes communities \cite{Nightscout,Tidepool}, primarily patients with Type 1 Diabetes. Table \ref{tab:dataset} provides an overview of two unique CGM datasets analyzed in this work. Dataset-1 includes 60 days of recordings from 10 subjects with diabetes while dataset-2 includes 100 - 270 days of recordings from 34 subjects with diabetes. There was no overlap between subjects across both datasets. As shown in Table \ref{tab:dataset}, there was a fair split of well-controlled (52\%) vs. poorly-controlled (48\%) subjects with diabetes based on the ADA's recommendation to maintain hemoglobin A1C < 7\% (equivalent to an average BG < 154 mg/dL) \cite{ADA2017,nathan2008translating}. Figure \ref{fig:subjBGCounts_5Cat} presents stacked bar plots showing subject-level BG distributions across the five notable categories. In dataset-1, subjects 1, 4, 8, 9, and 10 are examples of persons with well-controlled diabetes (i.e. average BG < 154 mg/dL or estimated A1C < 7\%). Meanwhile, in dataset-2, subjects 5, 11, 16, and 24 are examples of persons with poorly-controlled diabetes (i.e. average BG > 154mg/dL or estimated A1C > 7\%).

\begin{table*}
  \centering
  \caption{CGM dataset description. The values in parenthesis represent a breakdown of the number between well-controlled and poorly-controlled subjects with diabetes based on the ADA glycemic target criteria \cite{ADA2017}.} \label{tab:dataset}
  \begin{tabular}{c c c c c}
    \hline
    Dataset & Subjects & Ages & Days/Subject & Total Samples\\
    \hline
    1 & 10 (6,4) & unknown & 60 & 152,477\\
    2 & 34 (17,17) & 24 - 52 yrs. & 100 - 270 & 1,513,398\\
    \hline
  \end{tabular}
\end{table*}

%An example of CGM data can be found in a manufacturer's device report such as in \cite{Medtronic}.

\begin{figure}
    \centering
    \includegraphics[width=0.95\columnwidth]{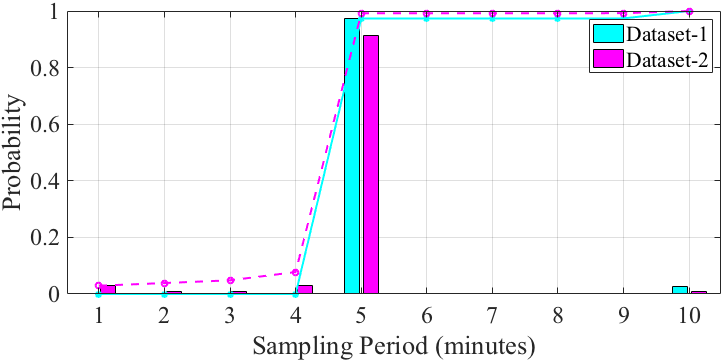}
    \caption{Probability Density Function (PDF) and Cumulative Distribution Function (CDF) of sampling periods of blood glucose readings in dataset-1 and dataset-2.}
    \label{fig:samplingPeriod}
\end{figure}

As part of the data-cleaning step, we removed duplicate, incomplete, and invalid samples. A valid data sample is one that includes a date, timestamp, and glucose reading in the range of 40 - 400 mg/dL. The data-cleaning step reduced dataset-1 and dataset-2 by 4.28\% and 18.89\%, yielding 152,477 and 1,513,398 samples, respectively. Based on today's technology, CGMs record a glucose value approximately every 5 minutes with the highest sampling rate being one sample every 1 minute upon the user's request \cite{desalvo2013continuous, fonda2013minding}. Figure \ref{fig:samplingPeriod} shows a probability density function of the sampling period and confirms that approximately 99\% of our dataset was recorded every 5 minutes with a less than 1\% sampled every 1 - 4 minutes. Given that a CGM is a wearable PHD, the user decides if and when to wear it. Therefore, missing data is not uncommon. The rest of our analysis investigates CGM adherence and influential factors that are explainable from the dataset.

\section{Analysis}\label{analysis}
Our  focus is to understand whether and to what extent management outcomes (i.e. achieving target goals) affect wearing behavior of CGMs. Toward this goal, we: 
\begin{enumerate}
    \item Investigate CGM wear time and explore sample distribution across the five key BG categories discussed in the "\nameref{background}" section.
    \item Characterize periods of missing data (known as \textit{ data gaps} in this work) by duration and distribution in the relevant BG categories.
    \item Perform statistical tests (i.e. One-way ANOVA and Two-Sample T-tests) to evaluate the significance of data gap durations in different BG categories.
    \item Investigate the duration of data gaps in normal vs. abnormal BG categories on subject- and group-levels, using subgroups based on management- and age-criteria.
\end{enumerate}
 
 %We analyzed the two diabetes datasets introduced in Section~\ref{dataset} to investigate users' adherence to CGM devices. present an analysis of distribution of gaps and their duration across the BG categories. 
 
\subsection{CGM Wear Time}

\begin{figure*}
\centering
\begin{subfigure}{0.49\textwidth}
    \centering
     \includegraphics[width=1.0\linewidth]{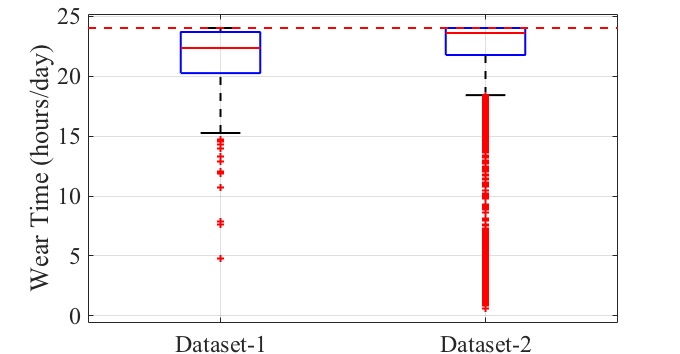}
    \caption{CGM wear time}
    \label{fig:wearingTime_ds1}
\end{subfigure}
%\newline
%\newline
\begin{subfigure}{0.49\textwidth}
    \centering
    \includegraphics[width=1.0\linewidth]{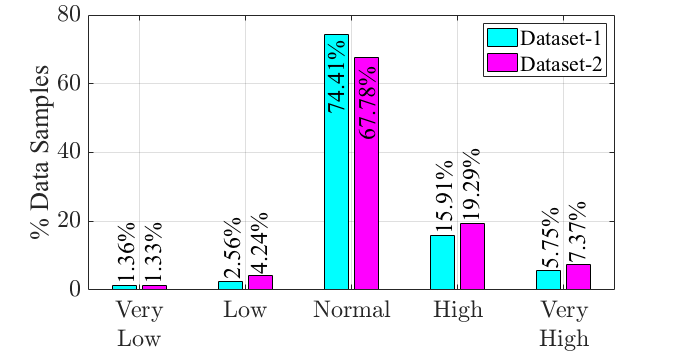}
    \caption{BG sample distribution}
    \label{fig:sampleCounts_5bg}      
\end{subfigure}
\caption{Overview of CGM wear time and sample distribution across the 5 key BG categories in both datasets.}
\label{fig:CGM_wearTime_and_SampleDistribution}
\end{figure*}

As defined in prior work \cite{tang2018adherence,meyer2017identification}, wear time is a count of the number of hours in a day that a PHD was worn. In this study, missing data was used as a proxy for calculating wear time of CGMs given that there will exist a recorded BG sample whenever the device is worn and turned-on for use. Figure~\ref{fig:wearingTime_ds1} presents an overview of wear time as determined by the presence of missing data in both datasets. We observe that majority of the time users wore their CGM device for greater than 20 hrs/day. The average wear time was 21.59 ($\pm$ 2.69) hrs/day and 22.16 ($\pm$ 3.63) hrs/day for dataset-1 and dataset-2, respectively. This is indicative of a generally higher adherence to prescription PHDs compared to non-prescription PHDs such as physical activity trackers with an average wear time of ~ 10-hrs/day \cite{jeong2017smartwatch, meyer2017identification, lyons2015dumbwatch}. However, as shown in Figure \ref{fig:wearingTime_ds1} there are several cases in which a CGM user's wear time in a given day is low (e.g. less than 15 hrs/day which is below the 25-th percentile mark in both datasets). We tailored our analysis on understanding such cases and potential associations with the user's BG readings (i.e. management outcomes) prior to the start of a data gap. Contiguous streams of missing data is used as a proxy for non-adherence to CGMs.

\subsection{Sample Distribution in BG Categories}

Figure \ref{fig:sampleCounts_5bg} presents an aggregate of CGM sample distribution across the five key BG categories for all subjects in this study. The highest percentages of BG readings, 74.4\% in dataset-1 and 67.7\% in dataset-2, are in the normal (or target) range. This is representative of positive management outcomes (i.e. subjects are meeting their goals in diabetes care). However, \ignore{ up to} $\approx$ 27\% of the data samples are in the \textit{high} BG range, with 15.9\% in dataset-1 and 19.3\% in dataset-2 in the \textit{high} range (> 180 mg/dL), and 5.8\% in dataset-1 and  7.4\% in dataset-2 in the \textit{very high} (> 250 mg/dL). Conversely, \ignore{ up to} $\approx$ 5\% of the data samples are in the \textit{low} BG range, with about 2.6\% in dataset-1 and 4.2\% in dataset-2 in the \textit{low} range (< 70 mg/dL), and 1.3\% of both datasets in the \textit{very low} (< 54 mg/dL). As described, \textit{low} and \textit{high} BG are representative of suboptimal management outcomes (i.e. subjects are \textit{not} meeting their goal in diabetes care). Given that low BG is more dangerous in the near-term than high BG value \cite{danne2017international}, the data distribution in these categories shows that clinically-significant categories (\textit{very low} and \textit{very high}) occur less often.

\begin{comment}
Next, we investigate users' stay in different BG categories based on available BG data samples. In particular, we are interested to find how often users reach to extreme categories, i.e., \textit{very low} and \textit{very high}. In Figure~\ref{fig:sampleCounts_5bg}, we observe that around 25\% (dataset-1) and 33\% (dataset-2) times users stay beyond the Normal BG category (\nameref{background} section). In next sections we will investigate whether these abnormal BG readings have association with users' adherence to the CGM devices. 
\end{comment}

\subsection{Data Gaps}

\begin{comment}
\begin{figure}
    \centering
    \includegraphics[width=0.6\columnwidth]{figures/box_subjLevel_gap_duration_d1_d2.png}
    \caption{Boxplot of gap durations (in minutes).}
    \label{fig:gapDurations}
\end{figure}
\end{comment}

\begin{figure*}
    \centering
    \includegraphics[width=1.75\columnwidth]{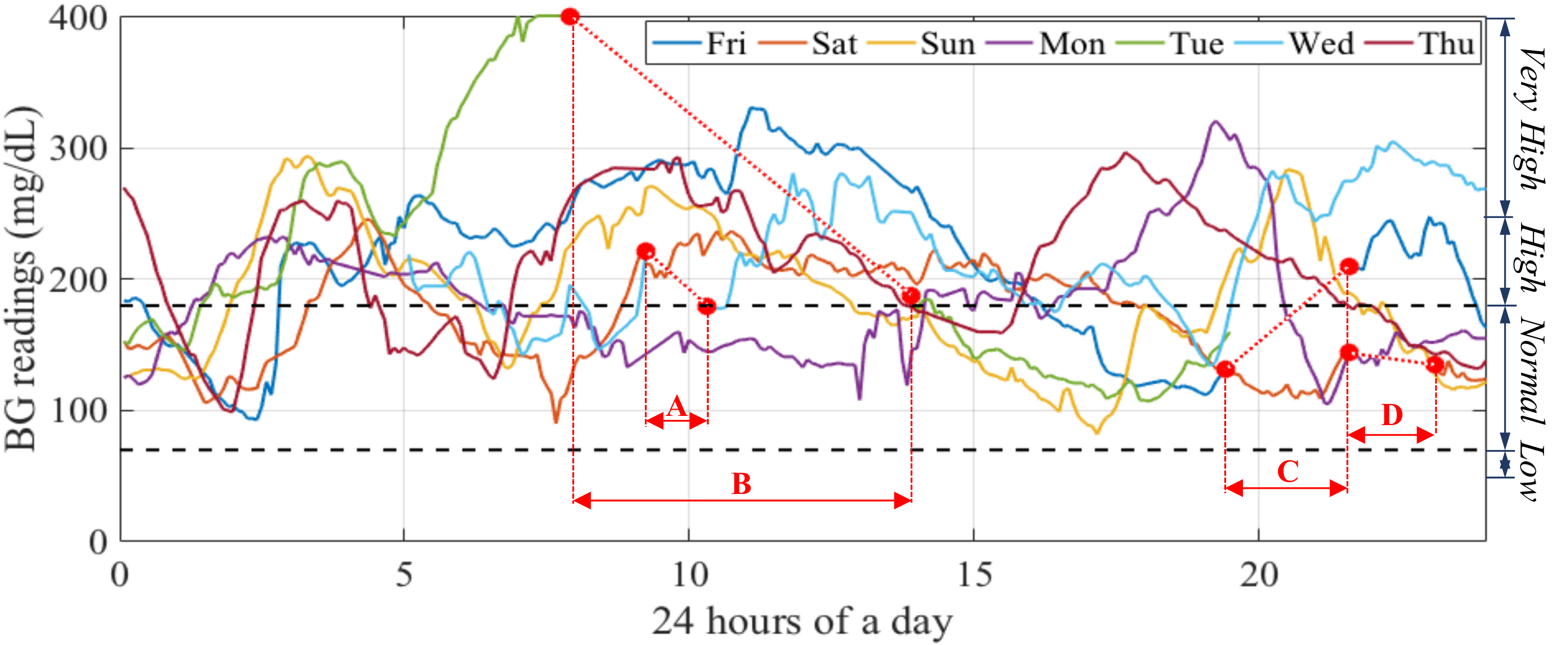}
    \caption{Example of CGM data from one subject in a single week - presented in the same format as a "daily overlay plot"\ignore{~\cite{Medtronic,DexcomClarity}}. The horizontal dashed lines indicate threshold values for \textit{low} and \textit{high} BG, respectively. Annotated segments show data gaps, where A, B, and D are data gaps associated with a decrease in BG, while C is a data gap associated with an increase in BG.}
    \label{fig:CBG_example}
\end{figure*}

Figure \ref{fig:CBG_example} shows a representative week of CGM data from one subject and highlights key concepts used in the remainder of this paper including data gaps and associations with an increase or decrease in BG. These concepts are further explained below: 

%our approach to define gaps and two possible cases of change in BG categories during the gaps. 
\begin{itemize}
\item  A {\bf data gap} ($\delta$) is a period in which there is no BG reading recorded on the continuous glucose monitor. This represents periods of contiguous missing data. In the ideal scenario, users should wear their prescription PHDs throughout the day (including at night-time) and the device will record BG data continuously at a preset sampling rate of 5-minutes. However, the sensor can malfunction or users may take the device off for different reasons, which can lead to missing data, i.e., a gap in the continuous recording. Informed by prior work on missing data and interpolation of CGM samples \cite{fonda2013minding}, a data gap is defined as: 

\begin{equation}
Data Gap = \delta \ge 2 \times mode(T) \wedge \delta < 24 \times 60
\end{equation}

where $\delta$ is the duration in minutes between adjacent data samples, T is the set of sampling periods in a day, and $mode(.)$ is the function used to find a number that occurs most often in a set of numbers. Therefore, a data gap is identified when there is missing data greater than twice the sampling period of 5-minutes (i.e. > 10-minutes) and within 24 hours.

\item An {\bf increase in BG} describes the scenario where a user's last BG reading before a data gap is lower than the BG reading after the gap. Based on the BG reading right before a string of missing data, we categorize data gap events into the five key categories discussed in the~\nameref{background} section. An increase in BG readings is most commonly influenced by food intake and may occur when a user is trending to or in a \textit{low} BG category \cite{desalvo2013continuous,danne2017international}. In our analysis, we investigate whether there is an association between the presence of data gaps immediately following \textit{low} or \textit{very low} BG readings. We also evaluate the length of data gaps in each BG category. This analysis aims to understand the influence of \textit{low} BG categories (i.e. suboptimal management) on non-adherence to CGM use in diabetes care. We seek to answer the question: are users more likely to take off their CGM during periods of \textit{low} BG and return to wearing their device when BG readings have increased (potentially back to the normal range)?

\item A {\bf decrease in BG} describes the scenario where a user's last BG reading before a data gap is higher than the BG reading after the gap. Based on the BG reading right before a string of missing data, we categorize data gap events into the five key categories discussed in the~\nameref{background} section. A decrease in BG readings is most commonly influenced by insulin use and may occur when a user is trending to or in a \textit{high} BG category \cite{desalvo2013continuous,danne2017international}. In our analysis, we investigate whether there is an association between the presence of data gaps immediately following \textit{high} or \textit{very high} BG readings. Likewise, we evaluate the length of data gaps in each BG category. This analysis aims to understand the influence of \textit{high} BG categories (also suboptimal) on non-adherence to CGM use in diabetes care. We seek to answer the question: are users more likely to take off their CGM during periods of \textit{high} BG and return to wearing the device when the BG readings have decreased (potentially back to the normal range)?
\end{itemize}

\section{Results}\label{results}
In this section, we evaluate wearing behavior of CGM devices in daily living, with a focus on: 1) non-adherence to CGMs, identified through data gaps (or missing data events), and 2) factors associated with non-adherence. It is important to note that unlike non-prescription PHDs such as FitBits, CGMs are prescription PHDs that should be worn throughout the day (including at night-time) to achieve optimal diabetes management. 

%\subsection{Non-adherence to CGMs}
\subsection{Distribution of Data Gaps in BG Categories}
\begin{figure*}
    \centering
    \includegraphics[width=1.75\columnwidth]{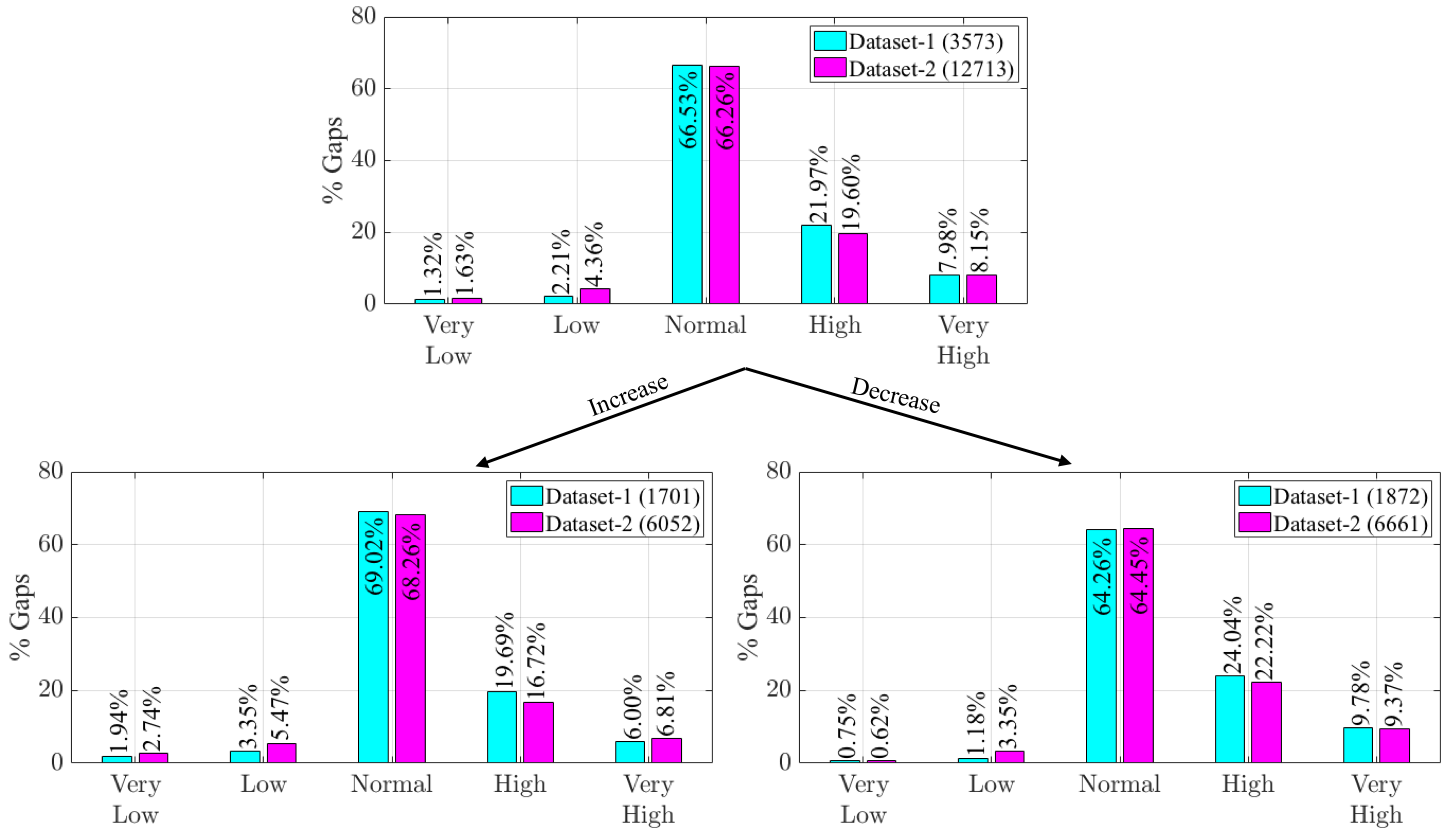}
    \caption{Data gap distribution in five key blood glucose categories (top) based on the last recorded sample and subcategories of BG increase (bottom left) and decrease (bottom right) immediately after the data gap. The parentheses in the legend show the total number of data gaps present in each dataset.}
    \label{fig:gapCounts_5bg}
\end{figure*}

For every data gap present, we evaluated the distribution of the last recorded sample in each of the five key BG categories. Figure \ref{fig:gapCounts_5bg} (top plot) shows that\ignore{ up to} $\approx$ 33\% of data gaps occurred when users were in abnormal BG categories (i.e. not achieving management goals). Furthermore, we investigated subcategories of increase (i.e. positive difference) and decrease (i.e. negative difference) in BG readings before and after the data gaps. This analysis aims to understand the distribution of data gaps for which the last recorded sample is \textit{very high} or \textit{high} and after the gap the first recorded sample is lower (i.e. a decrease in BG) - see segment B in Figure \ref{fig:CBG_example} for example. This could represent a scenario in which the user has an extreme reading, takes off their PHD, remedies the situation by taking medication, then returns to wearing the device after a while. We observe that data gaps associated with a decrease in BG reading have higher percentage of cases that start with \textit{high} and \textit{very high} (around 32\% in dataset-2 and 34\%  in dataset-1 -- Figure~\ref{fig:gapCounts_5bg} bottom right) compared to data gaps associated with an increase in BG reading (around 23\% in dataset-2 and 25\% in dataset-1 -- Figure~\ref{fig:gapCounts_5bg} bottom left). Similarly, for data gaps associated with an increase in BG reading, it is important to investigate the cases where the last recorded sample was a \textit{low} or \textit{very low} BG reading. We observe that data gaps with an increase in BG readings have higher percentages that start with \textit{low} and \textit{very low} (around 8\% in dataset-2 and 5\% in dataset-1 -- Figure~\ref{fig:gapCounts_5bg} bottom left) compared to data gaps associated with a decrease in BG value (around 4\% in dataset-2 and 2\% in dataset-1 -- Figure~\ref{fig:gapCounts_5bg} bottom right). 

The above analysis shows that there was a higher percentage of data gaps in the \textit{low} / \textit{very low} categories for which the BG value \textit{increased} immediately following the data gap. Similarly, there was a higher percentage of data gaps in the \textit{high} / \textit{very high} categories for which the BG value \textit{decreased} immediately following the data gap. This is suggestive of scenarios in which users took off their prescription PHD when \textit{not} achieving their goals and returned to wearing the PHD when their BG started trending toward the target goal. 

\begin{comment}
In Figure~\ref{fig:gapCounts_5bg}, we observe that around 33\% gaps starts with one of the four abnormal BG categories (Section~\ref{background}) with around 28\% (dataset-2) and 30\% (dataset-1) gaps start with \textit{high} or \textit{very high} BG category, which could potentially lead to insulin injection, followed by device take off based on users' level of comfort. Therefore, we further investigate the gap distributions for two sub-categories, where either BG readings increased (Figure~\ref{fig:gapCounts_5bg} bottom left) or decreased (Figure~\ref{fig:gapCounts_5bg} bottom right) after gaps. 

As discussed before for gaps with decreased BG readings it will be important to investigate the cases where users start with \textit{high} and \textit{very high} BG categories since these cases might be associated with insulin injection in order to drop BG category, followed by device take off. 

Similarly, for gaps with increased BG readings it will be important to investigate the cases where users start with \textit{low} and \textit{very low} BG categories since these cases might be associated with food intake in order to rise BG categories, followed by device take off. 
\end{comment}

%\subsection{Factors associated with non-adherence}
\subsection{Duration of Data Gaps in BG Categories}
\begin{figure*}
\centering
\begin{subfigure}{.5\textwidth}
  \centering
  \includegraphics[width=.95\linewidth]{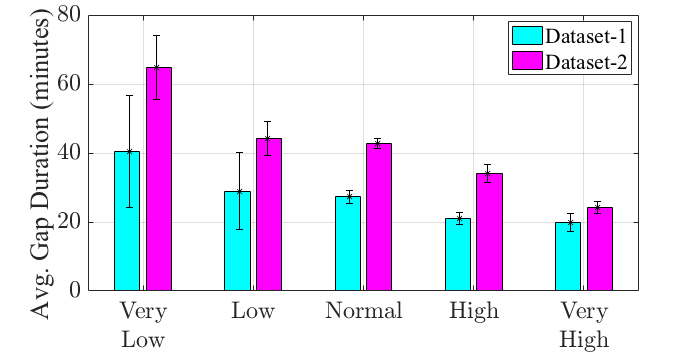}
  \caption{Data gaps associated with an increase in BG}
  \label{fig:gapDuration_Inc}
\end{subfigure}%
\begin{subfigure}{.5\textwidth}
  \centering
  \includegraphics[width=.95\linewidth]{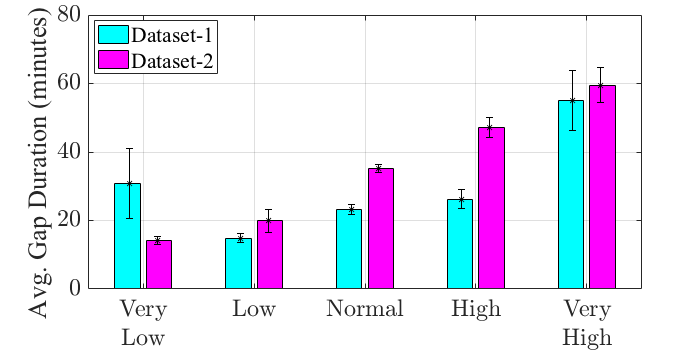}
  \caption{Data gaps associated with an decrease in BG}
  \label{fig:gapDuration_Dec}
\end{subfigure}%
\caption{Average gap durations with error bars across different blood glucose categories for cases when BG reading (a) increase and (b) decrease after gaps.}
\label{fig:gapDurations}
\end{figure*}

Figure~\ref{fig:gapDurations} presents an analysis of the duration of data gaps across each of the five key BG categories. On average, the length of missing data events ranges from 15 -- 70 minutes (mean = 34.21 minutes in dataset-1 and 48.79 minutes in dataset-2). A key observation is that the longest data gaps occurred in the most severe BG categories; equivalent to when users were farthest away from their target goal. More specifically, the \textit{very low} BG category has the longest data gap associated with an increase in BG after the gap -- Figure \ref{fig:gapDuration_Inc}. For this case, the duration of missing data in the \textit{very low} BG category is 1.5 times greater than the duration when users are in the \textit{normal} category. Conversely, the \textit{very high} BG category has the longest data gap associated with a decrease in BG value after the gap -- Figure \ref{fig:gapDuration_Dec}. Similarly, the duration of missing data when users were in the \textit{very high} categories is up to 1.5 times the duration when users were in the \textit{normal} category. This result is suggestive of scenarios in which CGM users take off their prescription PHD when they are in suboptimal BG categories (i.e. \textit{not} achieving their management goals). Additionally, users tend \textit{not} to wear the device for longer periods when the take-off started in a more severe or extreme BG category. It is also important to note that there is a similar trend between data gap duration and BG category across both independent datasets. This supports that the results observed are grounded and not biased to one specific dataset. 

% ($\pm$ 217.08) and ($\pm$ 222.29) 

\begin{comment}
Still there are several cases where the gap durations are longer than 1 hour. Therefore, we are interested to investigate those cases and their association with users' presence in different BG categories, especially extreme categories, i.e., \textit{very low} and \textit{very high}. 
\end{comment}

\begin{comment}
Next, we investigate gap durations for the cases when BG readings increase (Figure~\ref{fig:gapDuration_Inc}) and decrease (Figure~\ref{fig:gapDuration_Dec}) after the gaps. In Figure~\ref{fig:gapDuration_Inc} we observe that for gaps associated with increased BG readings have higher average durations for cases that start with \textit{low} and \textit{very low} BG categories. However, in Figure~\ref{fig:gapDuration_Dec} we observe an opposite pattern, i.e., for gaps associated with decreased BG readings have higher average durations for cases that start with \textit{high} and \textit{very high} BG categories. These findings show that users' take the device off for longer period when they reach to extreme BG categories, i.e., their adherence to CGM devices drop. 
\textcolor{red}{A quick possible remedy could be to make the CGM a bit more sensitive and proactive to send notifications to users before the BG readings cross the Normal categories.} In the next section, we will test the statistical significance of average gap durations among cases.
\end{comment}

\subsubsection{Statistical Significance of Data Gap Durations}

For significance testing, we use dataset-2 because it is larger and has more samples needed for a {\em One-way ANOVA} and {\em Two-Sample T-test} per the APA guidelines~\cite{apaGuidelines}. We first perform a {\em One-way ANOVA} test for the null hypothesis: ``the average duration of data gaps in different BG categories is the same.''
Table~\ref{ANOVA_ds2_inc_drop} shows the results which support to reject the null hypothesis with p-value $< 0.01$. Therefore, the average duration of data gaps starting in different BG categories is \textit{not} the same. Note that in this table, ``SS'', {\upshape``df''}, ``MS'', and ``F'' represent ``Sum of Squares'', ``degree of freedom'', ``Mean Square'', and ``F-statistic'',  respectively. $p-values$ are presents using three levels of $\alpha$, i.e., $p < .001$ (marked as ``***''), $p < .01$ (marked as ``**''), $p < .05$ (marked as ``*''), and $p > .05$ (marked as ``.''). 

%\textcolor{blue}{@Sudip: The write-up does not explain the results in Table 4. Also, I am wondering, is it necessary to still separate out "increase in BG" vs. decrease in BG for this one-way ANOVA and even the Two-sample T-test below? Based on the null hypothesis, I do not think there is a benefit doing two different statistical tests here. Let me know what you think. }

Next, to compare the average duration of data gaps in different BG categories, we perform a {\em Two-Sample T-test} with $H_0 : \mu_i=\mu_j$, where $\mu_i$ and $\mu_j$ represents the average data gap duration in two separate BG categories. For our comparison, we use the average gap duration of {\em normal} BG category as a reference and compare data gap durations in extreme BG categories, i.e., {\em very low} and {\em very high}. We also compare BG categories {\em low} vs. {\em very low} and {\em high} vs. {\em very high} to test for potential differences. 

Table~\ref{ttest_ds2_inc_drop} shows the {\em Two-Sample T-test} results.
We observe that the average gap duration in extreme BG categories is significantly different from the duration in the {\em normal} BG category. More specifically, $p-value = .0066$ for \textit{very low} vs. \textit{normal}, and $p-value = .0180$ for \textit{very high} vs \textit{normal}, respectively. Furthermore, we observe that the average gap duration starting in the {\em very low} BG category is significantly different from the average gap duration starting in the {\em low} BG category ($p-value = .0045$). Therefore, a {\em very low} BG category has a greater negative impact on users' adherence to the device compared to {\em low} BG category. On the other hand, the comparison of average gap durations in {\em very high} and {\em high} BG categories do not show a significant difference. This means {\em high} and {\em very high} BG categories could have a similar (not different) negative impact on users' adherence to the device. 

The above finding supports that duration of non-adherence to CGMs is significantly associated with severity of suboptimal management. Current CGMs have the ability to alarm users' when BG readings are trending toward out-of-target range (or abnormal) values. However, the utility of this feature is unknown and increased utility should be encouraged to improve adherence.

\begin{table}[b]
\begin{center}
\caption{{\em One-way ANOVA} table for testing the null hypothesis that "the average duration of data gaps in different BG categories is the same" - using dataset-2. The result shows to reject the null hypothesis.}
\begin{tabular}{l|c c c c c}
\hline
Source & SS & df &  MS & F & Prob$>$F \\
%\hline
\hline
Groups & 1.12e+05 & 4     & 2.81e+04 & 3.59 & 0.0062 \\
Error  & 9.93e+07 & 12708 & 7.81e+03 & & \\
Total  & 9.94e+07 & 12712 & & & \\
\hline
\end{tabular}
\label{ANOVA_ds2_inc_drop}
\end{center}
\end{table}

\begin{table}[b]
\begin{center}
\caption{{\em Two-Sample T-test} comparing the average data gap duration in different BG categories - using dataset-2. The last sample prior to a data gap was used to qualify BG categories. The results show statistically significant differences of data gaps in severe BG categories (very low/very high) vs. the normal BG category.}
\begin{tabular}{l|c|c|c}
\hline
Comparison  & df & {\em t-statistic} & significance \\
%\hline
\hline
\rowcolor{Gray} Very Low vs. Low        & 759 & 2.85 & ** \\
\rowcolor{Gray} Very Low vs. Normal     & 8629 & 2.72 & ** \\
\rowcolor{Gray} Very High vs. Normal  & 9458 & -2.37 & * \\
 Very High vs. High      & 3526 & -0.94 & . \\
\hline
\end{tabular}
\label{ttest_ds2_inc_drop}
\end{center}
\end{table}

\ignore{**************************************
\begin{figure}
    \centering
    \includegraphics[width=1.0\columnwidth]{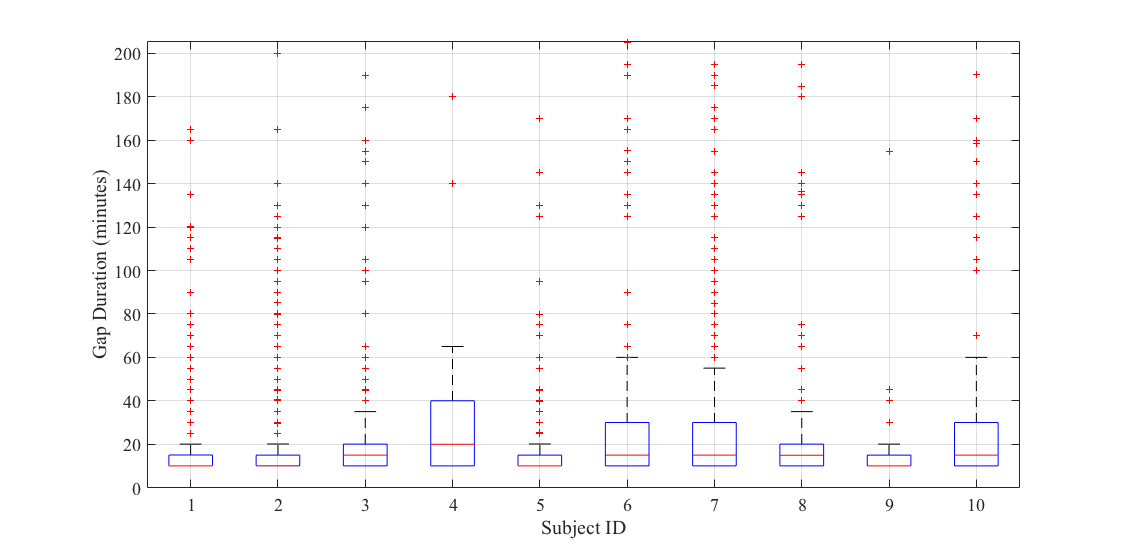}
    \caption{Box plots of subject-level data gap durations (dataset-1).}
    \label{fig:subjGapDurations_d1}
\end{figure}

\begin{figure*}
\centering
\begin{subfigure}{1.0\textwidth}
  \centering
  \includegraphics[width=.95\linewidth]{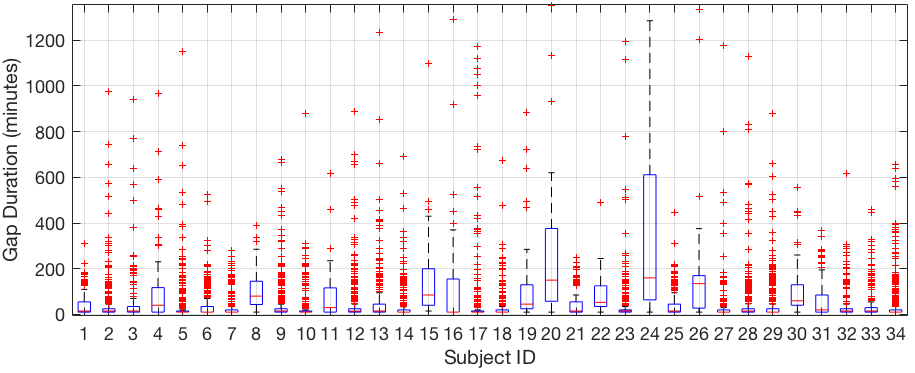}
  \caption{Original version}
  \label{fig:subjGapDurations_Ori}
\end{subfigure}%
\newline
\newline
\newline
\newline
\begin{subfigure}{1.0\textwidth}
  \centering
  \includegraphics[width=.95\linewidth]{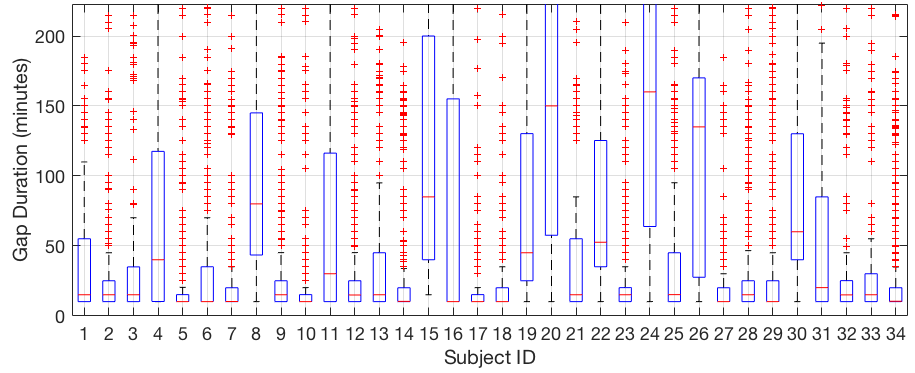}
  \caption{Zoomed/exploded version}
  \label{fig:subjGapDurations_zom}
\end{subfigure}%
\caption{Box plots of subject-level data gap durations (dataset-2).}
\label{fig:subjGapDurations_d2}
\end{figure*}
**************************************}

\begin{figure*}
\centering
    \begin{subfigure}{0.65\textwidth}
        \centering
        \includegraphics[trim=0 0 130 0mm,width=0.8\linewidth]{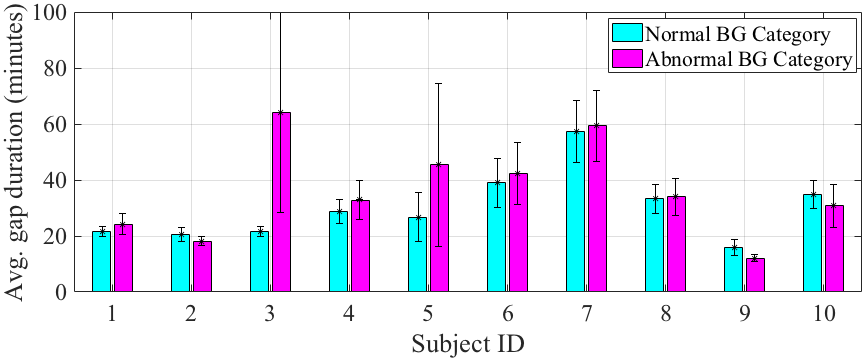}
        \caption{Average gap duration with error bars}
        \label{fig:bar_norm_abnorm_dt}
    \end{subfigure}%
    \newline
    \newline
    \newline
    %\newline
    \begin{subfigure}{0.65\textwidth}
        \centering
        \includegraphics[width=0.9\linewidth]{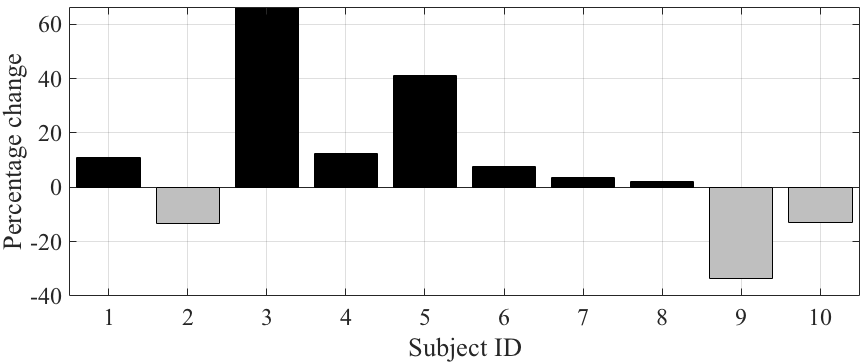}
        \caption{Percentage (\%) change in average gap durations}
        \label{fig:bar_norm_abnorm_dt_diff}
    \end{subfigure}%
\caption{Subject-level comparison of average gap duration in {\em normal} vs. {\em abnormal} BG categories for dataset-1.}
\label{fig:norm_abnorm_dt}
\end{figure*}

\subsection{Subject-Level Data Gap Analysis}

To further compare the average duration of data gaps during {\em normal} and {\em abnormal} (i.e. {\em very low}, {\em low}, {\em high}, and {\em very high}) BG categories, we calculate the difference in these values expressed as a percentage:
\begin{equation}
    \label{eq:percentChange}
    \% change = \frac{\bar{T}_{abnormal}-\bar{T}_{normal}}{\bar{T}_{abnormal}} \times 100
\end{equation}
Where $\bar{T}_{normal}$ and $\bar{T}_{abnormal}$ is average gap duration computed in the {\em normal} and {\em abnormal} BG categories, respectively. Using this equation, the $\% change$ can have one of the following values: 
\begin{subnumcases}{\% change=}
  > 0 & if $\bar{T}_{abnormal}>\bar{T}_{normal}$\label{case_pos}\\
  < 0 & if $\bar{T}_{abnormal}<\bar{T}_{normal}$\label{case_neg}\\
  0 & otherwise
  %0 & otherwise $\bar{T}_{abnormal}=\bar{T}_{normal}$\label{case_zero}
\end{subnumcases}

Figure \ref{fig:norm_abnorm_dt} shows the subject-level analysis of data gaps that started in the {\em normal} vs. {\em abnormal} BG categories - using dataset-1 as an example. Our analysis revealed that 70\% of subjects in dataset-1 and 50\% of subjects in dataset-2 had longer average gap durations that started in {\em abnormal} BG categories vs. {\em normal} BG category (i.e. case~\ref{case_pos}). This further supports the earlier finding that there exists an association between non-adherence to CGM use and suboptimal management (i.e. missing the target goal). It is important to note that this finding was more prevalent for some subjects (e.g. subjects 3 and 5) and not applicable to others (e.g. subjects 2 and 9). Therefore, it shows that this phenomenon is subject-dependent and \textit{not} generalizable across all people. This aligns with findings from prior work \cite{jeong2017smartwatch,meyer2017identification, doherty2017large} that factors which influence usage and adherence patterns to PHDs vary across individuals. The results of this paper add to this body of work by identifying missed health goals as a potential factor that contributes to non-adherence. 

\begin{table*}
\caption{{\em Two-Sample T-test} results for group-level data gap analysis. The result shows statistically significant differences in the average data gap duration of subgroups. Poorly-controlled subjects (A1C) > 7\% and older subjects (age > 40.38 yrs) had longer data gap durations (i.e. worse adherence to CGMs).}
\label{grp_ttest}
\begin{center}
\begin{tabular}{l|c|c|c|c|c|c}
\hline
Measure     & Grouping  & \multicolumn{2}{c|}{Avg. gap duration} &      &              & \\
\cline{3-4}
(threshold) &  criteria &  $\mu_1$ ($err$)  &  $\mu_2$ ($err$)  & df & {\em t-statistic} & significance \\
%\hline
%{\em Time in}  & $>=$,  & 41.30 &  49.39  & 17148  &  -2.37 & *\\
%{\em Range} (70\%)  &  $<$   &  (2.15)  & (2.49) &   &    &         \\
\hline
{\em Glycemic Target}  & $<=$,  & 40.01  & 51.56 & 17148 &  -3.42 & ***\\
{\em - A1C} (7\%)  &  $>$   & (1.69)  & (2.89) &   &    &         \\
\hline
{\em Age}  & $>=$,   & 41.91 & 37.42  & 13223  &  2.93  & ** \\ 
(40.38 yrs.)  & $<$   & (1.16) & (0.99)  &   &    &  \\ 
\hline
\end{tabular}
\end{center}
\end{table*}

\subsection{Group-Level Data Gap Analysis}

Per Table \ref{tab:dataset}, subjects in this study can be broken into subgroups to support the investigation of potential associations between distinct groups and non-adherence to PHDs. We performed a {\em Two-Sample T-test} with the null hypothesis: "the average duration of data gaps in different management- and age-subgroups is the same". This can be expressed mathematically as $H_0: \mu_1 = \mu_2$, where $\mu_1$ and $\mu_2$ is the average gap duration for each group. We used the ADA's glycemic target criteria of A1C less than 7\% ($\approx$ average BG < 154 mg/dL) \cite{ADA2017,nathan2008translating} as a threshold to divide subjects from both datasets into 2 subgroups: well-controlled (n = 23) vs. poorly controlled (n = 21) subjects with diabetes. Secondly, we used age as another criteria and the median age of 40.38 yrs as a threshold to divide subjects from dataset-2 into two equal-size groups (n=17): older vs. younger subjects.

Table~\ref{grp_ttest} shows the results from this analysis and supports to reject the null hypothesis that the average gap duration is the same across groups. We found that there was a statistically significant difference in the average gap duration ($p-value= .00063332$) of subjects with well-controlled diabetes vs. poorly-controlled diabetes. A key result is found is that subjects with poorly-controlled diabetes had a worse adherence to CGMs as evident through the missing data compared to subjects with well-controlled diabetes. This group-level analysis aligns with the earlier results that suboptimal outcomes (or missed target goals) is a potential factor that influences non-adherence to PHDs. Additionally, we found that older subjects had significantly ($p-value= .0034$) worse adherence to CGMs, as evident through more missing data, than younger subjects. This aligns with prior research \cite{doherty2017large,makela2013adherence} which identifies age to be a factor associated with varying adherence levels to PHDs, and even CGMs more specifically \cite{giani2017continuous}. These findings can guide tailored PHD design and interventions, although, it is important to note that there is individual heterogeneity as shown in Fig. \ref{fig:norm_abnorm_dt}, and the group-level finding is not a blanket statement for all people identified subgroups. 

%These findings will guide clinicians to better help diabetics patients to improve their CGM device adherence in order to better manage their health.

%In addition to {\em well-managed} and {\em not well-managed} grouping \ignore{based on {\em time in range} and {\em hemoglobin A1C} measures }discussed in Section~\ref{dataset}, we also consider subjects' age to divide the subject pool of dataset-2. We use the median age of 34 subjects, i.e. 40.38 years as the threshold.

\ignore{*******************************
\begin{table}[b]
\caption{{\em Two-Sample T-test} results for group-level data gap analysis. The result shows statistically significant differences in the average data gap duration of subgroups. Poorly-controlled subjects (A1C) > 7\% and older subjects (age > 40.38 yrs) had longer data gap durations (i.e. worse adherence to CGMs).}
\label{grp_ttest}
\begin{center}
\begin{tabular}{l|c|c|c|c|c|c}
\hline
%Measure     & Grouping  & \multicolumn{2}{c|}{Avg. gap duration} &      &              & \\
 Meas-    & Grou-  & \multicolumn{2}{c|}{Avg. gap} &      &              & sign- \\
 ure   & ping  & \multicolumn{2}{c|}{duration} &      &              & ifi- \\
\cline{3-4}
%(threshold) &  criteria &  $\mu_1$ ($err$)  &  $\mu_2$ ($err$)  & df & {\em t-statistic} & significance \\
(thres- &  crite- &  $\mu_1$  &  $\mu_2$  &  & {\em t-sta-} & can- \\
hold) & ria &  ($err$)  &  ($err$)  & df & {\em tistic} & ce \\
%\hline
%{\em Time in}  & $>=$,  & 41.30 &  49.39  & 17148  &  -2.37 & *\\
%{\em Range} (70\%)  &  $<$   &  (2.15)  & (2.49) &   &    &         \\
\hline
%{\em Glycemic Target}  & $<=$,  & 40.01  & 51.56 & 17148 &  -3.42 & ***\\
{\em Glycemic}  & $<=$,  & 40.01  & 51.56 & 17148 &  -3.42 & ***\\
{\em Target}  &   $>$   & (1.69)  & (2.89) &   &    &    \\
{\em - A1C}  &   &  &   &    &         \\
(7\%)  &     &   &  &   &    &         \\
\hline
{\em Age}  & $>=$,   & 41.91 & 37.42  & 13223  &  2.93  & ** \\ 
(40.38  & $<$   & (1.16) & (0.99)  &   &    &  \\ 
yrs.)  &        &  &   &    &         \\
\hline
\end{tabular}
\end{center}
\end{table}
*******************************}

\ignore{************************************
Age-based grouping: Use median value (40.38y) to split into subgroups
Senior citizens: 1 2 8 13 15 16 17 18 22 23 24 26 28 29 32 33 34
Younger citizens: 3 4 5 6 7 9 10 11 12 14 19 20 21 25 27 30 31

Potential Takeaway: Older subjects (i.e. > 40.38 yrs) have had worse adherent (in terms of avg. gap duration) to CGMs than younger subjects (i.e. <= 40.38 yrs).

avg_dt_3criteria_groupComparison_ds2
h = 1
p = 0.0034
ci = 1.4841    7.4919

stats = 
  struct with fields:

    tstat: 2.9286
       df: 13223
       sd: 88.1118
   
Following 2 analysis are for DS1 + DS2    
Measure: Time in Range ..........
mean = 41.3034   49.3868
err = 2.1525    2.4910
h = 1
p =  0.0177
ci = -14.7639   -1.4029
stats = 
    tstat: -2.3717
       df: 17148
       sd: 221.3074
       
Measure: Hemoglobin A1C .........
mean = 40.0068   51.5567
err = 1.6967    2.8944
h = 1
p = 6.3332e-04
ci = -18.1743   -4.9256
stats = 
    tstat: -3.4176
       df: 17148
       sd: 221.2683

A1C using only DS1
Measure: Hemoglobin A1C
mean = 22.7341   42.2950
err = 1.2430    5.9513
h = 1
p = 0.0047
ci = -33.1346   -5.9873
stats = 
    tstat: -2.8254
       df: 3882
       sd: 213.6720
       
A1C using only DS2
Measure: Hemoglobin A1C
mean = 44.2740   54.7105
err = 2.0903    3.3081
h = 1
p = 0.0071
ci = -18.0391   -2.8340
stats =
    tstat: -2.6908
       df: 13264
       sd: 223.3113
************************************}

\ignore{-------------------------------------
\begin{table}[b]
\begin{center}
\begin{tabular}{l|c c c c c}
\hline
Source & SS & df &  MS & F & Prob$>$F \\
%\hline
\hline
Groups & 8.51e+04 & 4    & 2.13e+04 & 5.88 & 1.03e-04 \\
Error  & 1.33e+07 & 3675 & 3.62e+03 & & \\
Total  & 1.34e+07 & 3679 & & & \\
\hline
\end{tabular}
\caption{{\em One-way ANOVA} table for gap durations across the five BG categories (dataset-1)}
\label{ANOVA_ds1_inc_drop}
\end{center}
\end{table}

\begin{table}[b]
\begin{center}
\begin{tabular}{l|c|c|c}
\hline
Comparison  & {\em df} & {\em t-statistic} & significance \\
%\hline
\hline
 Very Low vs. Low       & 127 & 0.84 & . \\
 Very Low vs. Normal    & 2521 & 1.36 & . \\
\rowcolor{Gray}  Very Low vs. High      & 839 & 1.67 & * \\
 Very Low vs. Very High & 335 & -0.41 & . \\
 Low vs. Normal         & 2550 & -0.02 & . \\
 Low vs. High           & 868 & 0.16 & . \\
 Low vs. Very High      & 364 & -1.47 & . \\
 Normal vs. High        & 3262 & 0.48 & . \\
\rowcolor{Gray}  Normal vs. Very High   & 2758 & -4.38 & *** \\
\rowcolor{Gray}  High vs. Very High     & 1076 & -4.07 & *** \\
 \hline
\end{tabular}
\caption{{\em Two-Sample T-test} results when comparing gap durations starting with different BG categories before the gaps (dataset-1)}
\label{ttest_ds1_inc_drop}
\end{center}
\end{table}

\begin{table}[b]
\begin{center}
\begin{tabular}{l|c c c c c}
\hline
Source & SS & df &  MS & F & Prob$>$F \\
%\hline
\hline
Groups & 2.71e+05 & 4    & 6.78e+04 & 9.21 & 2.03e-07 \\
Error  & 4.45e+07 & 6047 & 7.36e+03 & & \\
Total  & 4.46e+07 & 6051 & & & \\
\hline
\end{tabular}
\caption{{\em One-way ANOVA} table for gap durations, with increase in glucose categories after gaps, across the five blood glucose categories (dataset-2)}
\label{ANOVA_ds2_inc}
\end{center}
\end{table}

\begin{table}[b]
\begin{center}
\begin{tabular}{l|c|c|c}
\hline
Comparison  & {\em df} & {\em t-statistic} & significance \\
%\hline
\hline
\rowcolor{Gray} Very Low vs. Low & 495 & 2.13 & * \\
\rowcolor{Gray} Very Low vs. Normal & 4295 & 3.11 & ** \\
\rowcolor{Gray} Very Low vs. High & 1176 & 4.08 & *** \\
\rowcolor{Gray} Very Low vs. Very High & 576 & 6.29 & *** \\
Low vs. Normal & 4460 & 0.32 & . \\
Low vs. High & 1341 & 1.88 & * \\
Low vs. Very High & 741 & 4.17 & *** \\
Normal vs. High & 5141 & 2.79 & ** \\
Normal vs. Very High & 4541 & 4.23 & *** \\
High vs. Very High & 1422 & 2.34 & * \\
\hline
\end{tabular}
\caption{{\em Two-Sample T-test} results when comparing gap durations starting with different glucose categories before the gaps, followed by increase after gaps (dataset-2)}
\label{VL_ttest_ds2_inc}
\end{center}
\end{table}

\begin{table}[b]
\begin{center}
\begin{tabular}{l|c c c c c}
\hline
Source & SS & df &  MS & F & Prob$>$F \\
%\hline
\hline
Groups & 5.28e+05 & 4    & 1.32e+05 & 16.25 & 2.99e-13 \\
Error  & 5.41e+07 & 6656 & 8.13e+03 & & \\
Total  & 5.46e+07 & 6660 & & & \\
\hline
\end{tabular}
\caption{{\em One-way ANOVA} table for gap durations, with decrease in glucose categories after gaps, across the five blood glucose categories (dataset-2)}
\label{ANOVA_ds2_drop}
\end{center}
\end{table}

\begin{table}[b]
\begin{center}
\begin{tabular}{l|c|c|c}
\hline
Comparison  & {\em df} & {\em t-statistic} & significance \\
%\hline
\hline
Very Low vs. Low & 262 & -0.75 & . \\
Very Low vs. Normal & 4332 & -1.77 & * \\
\rowcolor{Gray} Very Low vs. High & 1519 & -1.88 & * \\
\rowcolor{Gray} Very Low vs. Very High & 663 & -2.27 & * \\
Low vs. Normal & 4514 & -2.95 & ** \\
\rowcolor{Gray} Low vs. High & 1701 & -3.56 & *** \\
\rowcolor{Gray} Low vs. Very High & 845 & -4.49 & *** \\
\rowcolor{Gray} Normal vs. High & 5771 & -4.61 & *** \\
\rowcolor{Gray} Normal vs. Very High & 4915 & -6.73 & *** \\
\rowcolor{Gray} High vs. Very High & 2102 & -2.19 & * \\
\hline
\end{tabular}
\caption{{\em Two-Sample T-test} results when comparing gap durations starting with different glucose categories before the gaps, followed by decrease after gaps (dataset-2)}
\label{VH_ttest_ds2_drop}
\end{center}
\end{table}

-------------------------------------}

\ignore{=============================
Dataset#1
tbl_inc_drop =
    {'Source'}    {'SS'        }    {'df'  }    {'MS'        }    {'F'       }    {'Prob>F'    }
    {'Groups'}    {[8.5067e+04]}    {[   4]}    {[2.1267e+04]}    {[  5.8777]}    {[1.0331e-04]}
    {'Error' }    {[1.3297e+07]}    {[3675]}    {[3.6182e+03]}    {0×0 double}    {0×0 double  }
    {'Total' }    {[1.3382e+07]}    {[3679]}    {0×0 double  }    {0×0 double}    {0×0 double  }

astats_inc_drop = 
    gnames: {5×1 cell}
         n: [50 79 2473 791 287]
    source: 'anova1'
     means: [36.3083 25.0042 25.1119 24.0203 42.1758]
        df: 3675
         s: 60.1517

Dataset#2
tbl_inc_drop =
    {'Source'}    {'SS'        }    {'df'   }    {'MS'        }    {'F'       }    {'Prob>F'  }
    {'Groups'}    {[1.1223e+05]}    {[    4]}    {[2.8056e+04]}    {[  3.5908]}    {[  0.0062]}
    {'Error' }    {[9.9291e+07]}    {[12708]}    {[7.8133e+03]}    {0×0 double}    {0×0 double}
    {'Total' }    {[9.9403e+07]}    {[12712]}    {0×0 double  }    {0×0 double}    {0×0 double}

astats_inc_drop = 
    gnames: {5×1 cell}
         n: [207 554 8424 2492 1036]
    source: 'anova1'
     means: [54.6792 34.4945 38.8407 41.8949 45.4333]
        df: 12708
         s: 88.3928

ap_inc =

   2.0290e-07

tbl_inc =

  4×6 cell array

  Columns 1 through 3

    {'Source'}    {'SS'        }    {'df'  }
    {'Groups'}    {[2.7112e+05]}    {[   4]}
    {'Error' }    {[4.4484e+07]}    {[6047]}
    {'Total' }    {[4.4755e+07]}    {[6051]}

  Columns 4 through 6

    {'MS'        }    {'F'       }    {'Prob>F'    }
    {[6.7779e+04]}    {[  9.2137]}    {[2.0290e-07]}
    {[7.3563e+03]}    {0×0 double}    {0×0 double  }
    {0×0 double  }    {0×0 double}    {0×0 double  }

astats_inc = 

  struct with fields:

    gnames: {5×1 cell}
         n: [166 331 4131 1012 412]
    source: 'anova1'
     means: [64.6887 44.2856 42.6906 34.1744 24.2218]
        df: 6047
         s: 85.7689

ap_drop =

   2.9902e-13

tbl_drop =

  4×6 cell array

  Columns 1 through 3

    {'Source'}    {'SS'        }    {'df'  }
    {'Groups'}    {[5.2844e+05]}    {[   4]}
    {'Error' }    {[5.4115e+07]}    {[6656]}
    {'Total' }    {[5.4644e+07]}    {[6660]}

  Columns 4 through 6

    {'MS'        }    {'F'       }    {'Prob>F'    }
    {[1.3211e+05]}    {[ 16.2490]}    {[2.9902e-13]}
    {[8.1303e+03]}    {0×0 double}    {0×0 double  }
    {0×0 double  }    {0×0 double}    {0×0 double  }

astats_drop = 

  struct with fields:

    gnames: {5×1 cell}
         n: [41 223 4293 1480 624]
    source: 'anova1'
     means: [14.1533 19.9614 35.1361 47.1741 59.4384]
        df: 6656
         s: 90.1682
=====================================}
\section{Discussion}\label{discussion}
In this study, we have investigated adherence to PHDs, with a focus on wearing behavior of CGMs used for diabetes management. We analyzed two independent datasets from a total of 44 subjects for 60 - 270 days and found that missing data (i.e. data gaps) is not uncommon. Our results show that suboptimal (i.e. \textit{low / high}) BG values is one factor that is associated with non-wearing behavior, identified through data gaps. Additionally, the length of data gaps is influenced by management outcomes, such that longer gap durations (i.e. periods of missing data) are significantly associated with extreme (i.e. \textit{very low / very high}) BG categories. It is important to note that the analysis in this work shows an association, not causality. Prior work supports that there are many reasons for data gaps in CGM readings, such as intermittent sensor error, sensor compression, and user errors \cite{fonda2013minding}. In addition to these, other factors associated with non-adherence to prescription PHDs include knowledge/education, age, associated costs, psychosocial, usability, and contextual factors \cite{desalvo2013continuous,makela2013adherence,rodbard2016continuous,giani2017continuous,raj2019BGHigher,raj2019clinical}. Nevertheless, the results of this paper highlight a critical dilemma. PHDs are developed to enable ubiquitous monitoring of health status, however, if users do not wear and use the devices consistently when not achieving the target goals then the benefit is limited. Conversely, if users wear and use the PHDs more often when they are achieving the target goals then the recorded data may be slightly biased and not a true reflection of the user's BG status. This is particularly important with regards to prescription PHDs, such as CGMs, given that doctors and care-givers rely on this information to understand and evaluate management and to guide treatment plans. 

%% New tests added by Sudip to discuss Temporal variation of gaps 

From our dataset-1, we found that majority of data gaps ($\approx$ 80\%) occurred during the daytime between the hours of 6AM and 12AM (i.e. midnight). Given that most users are likely to be awake and able to make wearing choices during the daytime, this observation is expected. However, we did not observe any significant differences between the average gap duration during the day versus at night. We also observed more data gaps during the weekends (i.e. Saturday and Sunday) compared to during the weekdays (i.e. Monday - Friday). But there were no significant differences between the average gap duration on weekdays vs. weekends. 
%This finding indicates that an hour during the day has a higher chance of observing gaps ($\approx$ 4.5\% gaps per hour) compared to an hour at night ($\approx$ 3.3\% gaps per hour), which might happen due to different distractions during the day time compared to the night when a user mostly sleeps. However, we do not observe any significant difference between the average gap duration during day and night. We also find that weekend days have a higher probability of observing gaps compared to weekdays, which might happen because of the differences among daily schedule. Similar to hours of a day, we do not find any clear pattern while comparing the average gap duration during weekend days and weekdays. These findings can be helpful to inform the caregivers and their clients, i.e., patients, to get the most benefit from the CGM.

To account for non-wearing behavior influenced by suboptimal management, we recommend that particular attention should be paid to the BG category users were in prior to the start of data gaps (or missing data events). This knowledge can be implemented in context-aware systems that include data-driven adherence analysis in embedded algorithms. Currently, CGM manufacturers such as Medtronic \cite{Medtronic} and Dexcom \cite{DexcomClarity} include "sensor wear (per week)" and "sensor usage" in their reports for patients, caregivers, and doctors. However, to the best of our knowledge, there is no analysis on when CGM devices are taken off. Therefore, if there is a pattern of users taking off their CGM device during periods of suboptimal management, this insight will be missed. Such adherence analysis is also applicable to other health domains in which PHDs are beneficial \cite{johnson2016methods}, especially as it relates to chronic disease management. For example, significant research has been committed toward wearable PHDs for continuous monitoring of blood pressure, stress, mental illness, and much more \cite{appelboom2014smart,bonato2010advances,chan2012smart}. As wearable PHD become a reality in other domains, adherence to these PHDs should be considered with specific attention paid to management outcomes when data gaps or missing data events occur. This analysis can inform targeted interventions to improve adherence to PHDs and health outcomes. Johnson et al. \cite{johnson2016methods} present other application spaces in which PHDs can serve dual-functions, namely for delivering medication and monitoring adherence to medical devices.

\subsection{Limitations}
Despite the interesting results found in this study, there are limitations that should be addressed in future work. First and foremost, given that the dataset was contributed by active members of online diabetes communities, these users are likely more invested in their health and may have better outcomes than the population at large. For example, Figure \ref{fig:wearingTime_ds1} shows the median wear time in both datasets is greater than 22-hours/day. This is relatively uncommon for prescription and non-prescription PHDs \cite{giani2017continuous,jeong2017smartwatch,meyer2017identification}. Additionally, the subject-inclusion criteria for this research was > 65\% wear-time for the range of data contributed. Therefore, a more general dataset will likely showcase worse suboptimal management and lower adherence to CGM or other PHDs. Nonetheless, as shown in Table \ref{tab:dataset}, the datasets in this study included representative samples from subjects with well-controlled and poorly-controlled diabetes based on the ADA glycemic target criteria \cite{ADA2017}, therefore, we expect that our results are reproducible.

Another limitation is the assumption that data gaps or missing data are directly indicative of non-adherence to PHDs, specifically CGMs in this case study. Majority of CGMs on the market today use a disposable sensor that has a lifetime of about 3 - 14 days depending on the device \cite{rodbard2016continuous}. Therefore, some data gaps are expected for sensor replacement and device restart. Additionally, some data gaps may be related to CGM battery replacement or recharge, although these are less likely since the battery life of CGM transmitters is about six months \cite{vashist2013continuous}. Future work will include follow-up interviews with users to understand reasons for data gaps and non-adherence to PHDs. This learning can further improve the design of such devices.

% Note: interventions to increase time in the normal range may increase wearability of devices (support for automated management)
 
\begin{comment}
 Thing to discuss: 
 - Difference between average wear time per day of physical activity trackers/smartwatches [12] vs. an average of about 22-hours per day in our dataset.
 
 Limitations:
 - An inclusion criteria our dataset was data sufficiency > 70\%, therefore, our dataset is more representative of persons with higher than average adherence to PHDs.
\end{comment}
\section{Conclusion and Future Work}\label{conclusion}

To the best of our knowledge, the work presented in this paper is one of the few studies that use quantitative, data-driven methods to understand day-to-day factors that affect adherence to prescription wearable medical devices. More specifically, our results suggest that adherence to PHDs is influenced by performance toward the target goal. With a focus on CGMs used in diabetes care, we found that $\approx$ 33\% of missing data occurred when users were not achieving their goal of maintaining BG within the \textit{normal} range. There was significantly longer durations of missing data when users were farthest away from the target goal (i.e. in extreme or more severe blood glucose categories). Additionally, subjects with poorly-controlled diabetes were observed to have significantly longer average data gap durations than subjects with well-controlled diabetes. This knowledge can inform the design of context-aware systems that include data-driven adherence analysis in embedded algorithms and provide interventions to improve outcomes.

As a starting point for future work, we recommend that PHD adherence analysis should combine qualitative evaluations of non-wearing behavior with data-driven analysis for a more comprehensive understanding of contributing factors. It is important to note that there should be a distinction between non-prescription PHDs and consumer wearable devices such as physical activity trackers and prescription PHDs such as CGMs. Given that PHDs for diabetes care are relatively advanced, this application space is ideal for learning insights that can influence future development and use of data from such devices. Future work following this study will explore other contextual factors that influence missing data events as well as good and/or suboptimal management in daily living. The long-term goal is to develop data-driven decision-support tools to improve health.
%\input{acknowledgments}

% REFERENCES FORMAT
% References must be the same font size as other body text.
\bibliographystyle{SIGCHI-Reference-Format}
\bibliography{references}

\end{document}